\documentstyle[aps,preprint]{revtex}
\begin{document}
\draft
\title{Corrections to scaling in 2--dimensional polymer statistics}
\author{S. R. Shannon, T. C. Choy and R. J. Fleming}
\address{Department of Physics, Monash University, Clayton,
Victoria, Australia}
\date{\today}
\maketitle
\begin{abstract}
Writing $\langle R^2_N \rangle = AN^{2\nu}(1+BN^{-\Delta_1}+CN^{-1}+ ...)$ for
the mean square end--to--end length $\langle R^2_N\rangle$\ of
a self--avoiding polymer chain
of $N$ links, we have calculated $\Delta_1$ for the two--dimensional {\em
continuum\/} case from a new {\em finite\/} perturbation method based on
the ground state of Edwards self consistent solution which predicts the
(exact) $\nu=3/4$ exponent. This calculation yields $\Delta_1=1/2$.
A finite size scaling analysis of
data generated for the continuum using a biased sampling Monte Carlo algorithm
supports this value, as does a re--analysis of exact data for
two--dimensional lattices.
\end{abstract}

A polymer chain is self--avoiding due to the excluded volume
effect between monomer units which causes an expansion
or `swelling' of the chain when compared to the free random walk.
The central quantity of interest is therefore the mean square
end--to--end length $\langle R^2_N\rangle$. This is believed to have the form
\begin{equation}
\langle R^2_N \rangle = AN^{2\nu}(1+BN^{-\Delta_1}+CN^{-1}+ ...) ,
\label{rform}
\end{equation}
where $N$ is the number of chain links, $\nu$ is the leading scaling exponent,
$A,B,C$ are excluded volume dependent coefficients and $\Delta_1$ is the
leading correction--to--scaling exponent.  It is now firmly
established  \cite{nienhuis,guttmann,ishinabe,santos} that in two
dimensions (2D) $\nu=3/4$ is exact. Despite this, there is very little
agreement
about the value of $\Delta_1$. Nienhuis \cite{nienhuis}
predicts $\Delta_1=3/2$, while Rapaport \cite{rapaport} has argued
that there is no need for a correction term other than the analytic
correction, i.e., $\Delta_1=1$. However, many numerical studies have disagreed
with these results, with estimates for $\Delta_1$ of $1.2$ \cite{havlin},
 $0.84$ \cite{lyklema} and $0.65$ \cite{ishinabe,santos,privman}.
These numerical estimates are based on results obtained from self--avoiding
walks on 2D lattices. With the exception of a very few authors
\cite{croxton,kremer} (these studies however were not concerned with the
correction to scaling terms) it appears that little work has been
done in the continuum.  Theoretical results are also in disagreement.
Besides Nienhuis's prediction, which relies on
a mapping to an exactly solvable solid--on--solid model on the
honeycomb lattice,
Baker~{\em et al\/} \cite{baker} predict \mbox{$\Delta_1=1.18$} using
RG arguments, while Saleur \cite{saleur} predicts
\mbox{$\Delta_1=11/16$} by conformal invariance. Interestingly, Saleur also
gives evidence for a term $\Delta_1=1/2$, but he then rejects this result.
Perturbation expansion techniques \cite{fixman,muthukumar}, which start from
the free random--walk solution, have also been used to predict
$\langle R^2_N\rangle$, but these methods have resulted in series
which are divergent in $N$ and $v$, the excluded volume parameter, and hence
a value for $\Delta_1$ cannot be predicted.
The obvious confusion in both the numerical and theoretical estimates
for $\Delta_1$, lack of corresponding data for the continuum, and the
possibility of using a better perturbation expansion
to determine $\langle R^2_N\rangle$\ form the motivation of this study.

We have used a new perturbation method,
which unlike previous studies, starts from a ground state that already
correctly predicts the {\em exact\/} large $N$ behaviour in 2D, namely
the Edwards self--consistent solution \cite{edwards}. Although it has
been shown \cite{fisher}
that the Edwards solution cannot be the correct form for the
self--avoiding random walk end--to--end distribution function \cite{bishop},
it has mathematically convenient features that enable a
perturbation expansion to be performed. We believe its use here underpins the
essential physics and that $\Delta_1$ thus obtained may well be
exact in 2D.
In path integral representation \cite{wiegel}, the exact distribution
function, or Green's
function, for the end--to--end distance ${\rm{\bf R}}$ is
\begin{equation}
G({\rm{\bf R}},L) = \int^{{\rm{\bf r}}(L)={\rm{\bf R}}}_{{\rm{\bf r}}(0)
={\rm{\bf 0}}}
D[{\rm{\bf r}}]\exp\left(-\frac{1}{l}\int^L_0ds
(\frac{\partial {\rm{\bf r}}(s)}{\partial s})^2 -
\frac{v}{l^2}\int^L_0ds\int^L_sds'\delta^2[{\rm{\bf r}}(s)-
{\rm{\bf r}}(s')]\right) ,
\end{equation}
where $L$ is the total chain length $L=Nl$, $l$ being the step length of one
link, and $v$ is the excluded volume. Two problems
arise in dealing with this intractable path integral. Firstly,
divergences appear in the calculation which must be handled carefully, and
secondly, the resulting series expansion is a power series of increasing $L$
and $v$. This leads to a divergent result unless the value of $v$ is
assumed to be very small. This divergent property is the hallmark of modern
critical phenomena theory whose resolution was
offered by the renormalization group
approach \cite{skma,amit}. Historically Edwards avoided
the divergence problems of such an approach by replacing
the point contact potential by a self--consistent field $W(r)$ which
in 2D is equal to $v\tilde{p}(r)/l$,\, where $\tilde{p}(r)$ is the
one--particle potential proportional to $r^{-2/3}$ \cite{edwards,myself}.
Therefore
\begin{equation}
W(r) = {\cal C}v^{2/3}r^{-2/3} ,
\end{equation}
where ${\cal C}=(\sqrt{3}/4\pi l)^{2/3}$. Thus the Edwards Green's function
$G_E({\rm{\bf R}},L)$ becomes
\begin{equation}
G_E({\rm{\bf R}},L)=\int^{{\rm{\bf r}}(L)={\rm{\bf R}}}_{{\rm{\bf r}}(0)
={\rm{\bf 0}}}D[{\rm{\bf r}}]
\exp\left(-\frac{1}{l}\int^L_0ds\left(\frac{\partial {\rm{\bf r}}(s)}{\partial
s}\right)^2 -
\int^L_0 W(s) ds \right) .
\label{ed_sol}
\end{equation}
Our approach relies on obtaining a better
first order perturbation expansion by starting from the Edwards ground state
and then perturbing this with the {\em difference\/} between the
self--consistent field and the true point contact potential \cite{ourmethod}.
Thus
\begin{eqnarray}
G({\rm{\bf R}},L) &=&
\int^{{\rm{\bf r}}(L)={\rm{\bf R}}}_{{\rm{\bf r}}(0)={\rm{\bf 0}}}
D[r]\exp\left(-\frac{1}{l}\int^L_0ds\left(\frac{\partial {\rm{\bf r}}(s)}
{\partial s}\right)^2 -
\int^L_0 W(s) ds \right) \nonumber \\
& &\times \exp\left( \int^L_0 W(s) ds -
\frac{v}{l^2}\int^L_0ds\int^L_sds'\delta^2[{\rm{\bf r}}(s)-
{\rm{\bf r}}(s')]\right) ,
\label{solution}
\end{eqnarray}
where the difference potential in the second exponential
term is now being treated as a perturbation.
The Fourier transform of Eq.\ (\ref{solution}) thus becomes
\begin{equation}
\hat{G}({\rm{\bf k}},L) = \hat{G}_E({\rm{\bf k}},L) + \hat{G}_1({\rm{\bf k}},L)
+ \hat{G}_2({\rm{\bf k}},L) + ... ,
\label{series}
\end{equation}
where $\hat{G}_1({\rm{\bf k}},L)$ and $\hat{G}_2({\rm{\bf k}},L)$ are the first
order terms in the perturbation expansion.
By using the method of Fourier and Laplace transformation
as in \cite{muthukumar}, we derive the following functions to leading order
in~$L$
\begin{eqnarray}
\hat{G}_E({\rm{\bf k}},L) & = & 1-\frac{k^2}{4}({\cal A}^2L^{3/2} +
\frac{3}{2{\cal B}}L + ...) , \nonumber \\
\hat{G}_1({\rm{\bf k}},L) & = &
\frac{3}{16}k^2 {\cal A}^2 L^{3/2} + ... , \label{g_vals} \\
\hat{G}_2({\rm{\bf k}},L) & = & 2L\Phi-2\Psi
-k^2(\frac{2{\cal A}^2\Phi}{5}L^{5/2} + \frac{3\Phi}{4{\cal B}}L^2 -
{\cal A}\Psi L^{3/2} + ... ) , \nonumber
\end{eqnarray}
where ${\cal A}$ and ${\cal B}$ are excluded volume and step length dependent
quantities that appear in the 2D Edwards solution, and $\Phi$
and $\Psi$, which also depend on ${\cal A}$ and ${\cal B}$,
 are well behaved convergent integrals in the large $L$ limit \cite{myself}.
The calculation of these functions required the exact form of the $L$
dependent normaliztion factor of the 2D Edwards solution. However, unlike
the normalization for the free walk solution which leads
to logarithmic divergences and hence the introduction of a cutoff $\epsilon$,
see \cite{muthukumar}, the normalization for the
2D Edwards solution has a form such that a term corresponding to $\epsilon$
appears naturally. As we know its exact $L$ dependence, integrals which
would otherwise diverge remain controlled. The subsequent calculation
of $\langle R^2\rangle$\ from these results gives
\begin{equation}
\langle R^2 \rangle = \frac{4}{5}{\cal A}^2L^{3/2}(1+\frac{15}{8{\cal A}^2
{\cal B}}L^{-1/2} -
\left(\frac{5}{4}\frac{16w\Psi+1}{8w\Phi}\right)L^{-1} + ... ) .
\label{final}
\end{equation}
It should be noted that in deriving this result, unlike previous
perturbation calculations, no divergences are encountered,
and no restriction is placed on the value of
$v$ since the series is convergent in $L$.
When comparing Eq.\ (\ref{final}) with Eq.\ (\ref{rform}), we see that
it predicts a value $\Delta_1=1/2$, as well as a {\em negative\/} value of
the coefficient~$C$. If we assume $a/l=0.5$, corresponding
to the maximum excluded volume $v=\pi a^2$, we calculate
${\cal A}\approx 0.793$, ${\cal B}\approx 1.13$, $\Psi\approx 0.107$
and $\Phi\approx 0.226$. Substituting these into Eq.~(\ref{final}) the
scaling amplitudes of Eq.~(\ref{rform}) become
$A\approx0.50$, $B\approx2.65$ and
$C\approx-2.07$. However these `mean field' values for the amplitudes are not
expected to agree well with numerical or exact results \cite{cardy}.

We now turn to numerical studies. In order to create 2D self--avoiding
chains in the continuum we have used
a {\em biased\/} sampling Monte Carlo method dating back
to Rosenbluth and Rosenbluth \cite{rosenbluth}. Although more efficient
algorithms exist for creating longer chains \cite{binder} we are
unsure about their reliability for studying the correction-to-scaling terms
in the continuum.  As our chains are in the continuum, the simulation
procedure is considerably more complicated than that in \cite{rosenbluth}.
Given a chain consisting of $n$
circles, the area available to the $(n+1)$th circle must be determined. If no
areas are large enough, the chain is discarded and a new one started, otherwise
the position of the next circle is picked randomly from the available areas
and the chain at that position weighted with a factor
$\theta/(2\pi\!-\!\beta)$, where $\theta$ is the total
available angle and $\beta$ is the angle excluded by the $(n\!-\!1)$th circle
as no doubling back is permitted. This weighting factor distinguishes
polymer statistics from the `true' self--avoiding walk \cite{parisi}.
In this way relatively long chains can easily be built even with
the maximum excluded volume. We have checked the program
against the analytic solution for the three step walk \cite{integrals}
and against a similar simple sampling program up to much longer
chain lengths, with satisfactory results. The method has also been tested
against Guttman's exact ennumeration data \cite{guttmann}.

To analyse our data, we have used the finite size scaling method
of Privman and Fisher \cite{privandfish} which is based on the
cancellation of leading terms. We plot the estimating function
\begin{equation}
A_{N,k}(\Delta) = \frac{N^{\Delta-2\nu}R^2_N-(N-k)^{\Delta-2\nu}R^2_{N-k}}
{N^\Delta-(N-k)^\Delta} .
\label{acalc}
\end{equation}
Assuming Eq.\ (\ref{rform}) with $\nu=3/4$, then
\begin{equation}
A_{N,1}(\Delta) = A + (1-1/\Delta_1)ACN^{-1}+ ... ,
\label{adelta}
\end{equation}
when $\Delta=\Delta_1$. The curves for different $N$ will thus cross at
a point close to the correct value of $A$ and $\Delta_1$ {\em assuming $|C|$
is small compared to $|A|$}. The insert of Fig. \ref{fig1} shows
this technique applied to the exact square lattice data of
Guttmann \cite{guttmann} for values $N=15$ to $N=27$. In this case $k=2$
is used and the resulting $A_{N,2}(\Delta)$ data averaged
with previous $A_{N-1,2}(\Delta)$ result to eliminate the
odd--even effect.  As can be seen the
curves cross at a value $\Delta\approx0.65$ and $A\approx0.765$,
in agreement with Ishinabe \cite{ishinabe}, while $\Delta_1=0.65$ was
also reported by Privman \cite{privman} using the same technique on the
triangular lattice data of Grassberger \cite{grassberger}. However,
these authors assumed $|C|\!\ll\!|A|$, even though no clear evidence was given
to support this assumption.  Figure \ref{fig1} also shows the results obtained
using the same method on data calculated from Eq.\ (\ref{rform}) (hereafter
called `simulated data') with $A=0.760,\ B=0.227,\ C=0.18$ and
$\Delta_1=0.5$. These values of $A$, $B$ and $C$ were obtained from a least
squares fit to the data \cite{guttmann} having set $\Delta_1=0.5$.
The resulting curves are virtually indistinguishable from
the exact data even though the values of $\Delta_1$ differ. We have confirmed
that any positive value of $C$ will shift the crossing
point to higher values of $\Delta$.   It thus appears that
for the lattice $\Delta_1$ could be as large as $0.66$, but
$\Delta_1=0.5$ is possible. The same procedure using
exact triangular lattice data \cite{grassberger} yielded excellent agreement
with simulated data for $A=0.704$, $B=0.175$, $C=0.128$ and $\Delta_1=0.5$.

Figure\ \ref{fig2} shows the resulting curves
when the estimating function Eq.\ (\ref{acalc}) is
applied to our continuum data for chain lengths of $N=10$ to $N=25$
with the maximum excluded volume ratio of~$0.5$. The estimate
of $\langle R^2_N\rangle$\ for $N=25$ comes from averaging approximately
$1.87\times 10^8$ walks, resulting in an error of less than $0.1\%$.
Clearly the data show no sign of crossing, and we suggest that this is due to
a large negative $C$. An estimator, similar to Eq.\ (\ref{acalc}),
but which gives simultaneous estimates of $B$ and $\Delta_1$ \cite{ishinabe}
was also studied, but it too showed no evidence of crossing. Due to the large
$C$ value and small random errors in our Monte Carlo continuum data
it is difficult to use other graphical techniques \cite{santos,lyklema} to
determine $\Delta_1$. We therefore use the above method of comparison
between simulated and Monte Carlo data.

We first assumed a value of $\Delta_1=0.666$, in agreement with \cite{santos},
and using a least squares curve fit to our data from $N=10$ to $N=25$,
we found a best fit with coefficient values of $A=1.023,\ B=0.491$
and $C=-0.681$. We then used these values to create simulated
$\langle R^2_N\rangle$\ data
to which we applied the $A_{N,1}(\Delta)$ vs. $\Delta$ analysis. The
resulting curves are shown in Fig.\ \ref{fig3}. They are very different
from those of the Monte Carlo continuum data. We then assumed a value
of $\Delta_1=0.5$ from which we obtained a best fit with coefficient values of
$A=0.990,\ B=0.489$ and $C=-0.839$. The resulting curves,
as shown in Fig. \ref{fig4}, are in excellent agreement with the
continuum data. The large negative $C$ should be noted. Thus to
assume $C$ is negligible, as has often been done
when analysing 2D lattice data \cite{ishinabe,santos,privman},
is a mistake and could lead to poor estimates
of $\Delta_1$.

In conclusion, we have presented strong evidence
that $\Delta_1=1/2$ for 2D chains in the
continuum. Although this value does not agree with any of those suggested
by other authors for 2D chains on the lattice
\cite{ishinabe,santos,lyklema,privman}, their data are compatible with
$\Delta_1=0.5$ when the effects due to the next order term in Eq.\
(\ref{rform})
are considered. Unless there is a breakdown
of the universality of both $\nu$ and $\Delta_1$
\cite{amit,barma}, we suggest that Saleur's \cite{saleur}
rejection of $\Delta_1=1/2$ should be re--examined.
Since our new perturbation method and our Monte Carlo analysis agree, we
suggest that the leading correction--to--scaling term in two--dimensions
is $\Delta_1=1/2$.

TCC would like to thank Prof. Sir S. F. Edwards, Prof. D. Sherrington,
Prof. R. Stinchcombe and Prof. M. Barma for helpful discussions. SRS
acknowledges the support of an Australian Postgraduate Research Award.

\begin{figure}
\caption{Plot of $A_{N,2}(\Delta)$ vs. $\Delta$ for simulated
square lattice data with
$A=0.760$, $B=0.227$, $C=0.18$ and $\Delta_1=0.5$. The insert shows the same
plot for the exact square lattice data of {\protect\cite{guttmann}}.}
\label{fig1}
\end{figure}

\begin{figure}
\caption{Plot of $A_{N,1}(\Delta)$ vs. $\Delta$ for our Monte Carlo 2D
continuum data.}
\label{fig2}
\end{figure}

\begin{figure}
\caption{Plot of $A_{N,1}(\Delta)$ vs. $\Delta$ for simulated
continuum data with
$A=1.023$, $B=0.491$, $C=-0.681$ and $\Delta_1=0.666$.}
\label{fig3}
\end{figure}

\begin{figure}
\caption{Plot of $A_{N,1}(\Delta)$ vs. $\Delta$ for simulated
continuum data with
$A=0.990$, $B=0.489$, $C=-0.839$ and $\Delta_1=0.5$.}
\label{fig4}
\end{figure}


\begin{references}
\bibitem{nienhuis} B. Nienhuis, Phys. Rev. Lett. {\bf 49}, 1062 (1982).
\bibitem{guttmann} A. J. Guttmann, J. Phys. A {\bf 20}, 1839 (1987).
\bibitem{ishinabe} T. Ishinabe, Phys. Rev. B {\bf 39}, 9486 (1989).
\bibitem{santos} Z. V. Djordjevic, I. Majid, H. E. Stanley and R. J. dos
Santos, J. Phys. A {\bf 16}, L519 (1983).
\bibitem{rapaport} D. C. Rapaport, J. Phys. A {\bf 18}, L39 (1985).
\bibitem{havlin} S. Havlin and D. Ben-Avraham, Phys. Rev. A {\bf 27}, 2759
(1983).
\bibitem{lyklema} J. W. Lyklema and K. Kremer, Phys. Rev. B {\bf 31},
3182 (1985).
\bibitem{privman} V. Privman, Physica A {\bf 123}, 428 (1984).
\bibitem{croxton} C. A. Croxton, J. Phys. A {\bf 19}, 987 (1986).
\bibitem{kremer} K. Kremer, A. Baumg\"{a}rtner and K. Binder, Z. Physik B
{\bf 40}, 331 (1981).
\bibitem{baker} G. A. Baker, Jr., B. G. Nickel, M. S. Green and D. I. Meiron,
Phys. Rev. Lett. {\bf 36}, 1351 (1976); Phys. Rev. B {\bf 17}, 1365 (1978).
\bibitem{saleur} H. Saleur, J. Phys. A {\bf 20}, 455 (1987).
\bibitem{fixman} M. Fixman, J. Chem. Phys. {\bf 23}, 1656 (1955).
\bibitem{muthukumar} M. Muthukumar and B. G. Nickel, J. Chem. Phys. {\bf 80},
5839 (1984).
\bibitem{edwards} S. F. Edwards, Proc. Phys. Soc. {\bf 85}, 613 (1965).
\bibitem{fisher} M. Fisher, J. Chem. Phys. {\bf 44}, 616 (1966). J. des
Cloizeaux, J. Physique (Paris) {\bf 31}, 715 (1970).
\bibitem{bishop} M. Bishop and J. H. R. Clarke, J. Chem. Phys. {\bf 94}, 3936
(1991).
\bibitem{wiegel} F. W. Wiegel, {\em An Introduction to Path--Integral Methods
in Physics and Polymer Science\/} (World Scientific, Singapore, 1986), p 44.
\bibitem{skma} S. K. Ma, {\em Modern Theory of Critical Phenomena\/} (Benjamin,
Massachusetts, 1976), p 116.
\bibitem{amit} D. J. Amit, {\em Field Theory, the Renormalization Group and
Critical Phenomena\/} (World Scientific, Singapore, 1984), p 189.
\bibitem{myself} S. R. Shannon and T. C. Choy. Unpublished.
\bibitem{ourmethod} This approach is new and appears to have been
missed in the literature for 30 years.
\bibitem{cardy} J. L. Cardy and A. J. Guttmann, J. Phys. A {\bf 26}, 2485
(1993).
\bibitem{rosenbluth} M. N. Rosenbluth and A. W. Rosenbluth, J. Chem. Phys.
{\bf 23}, 356 (1954).
\bibitem{binder} A. Baumg\"{a}rtner and K. Binder, J. Chem. Phys. {\bf 71},
2541 (1979).
\bibitem{parisi} D. J. Amit, G. Parisi and L. Peliti, Phys. Rev. B {\bf 27},
1635 (1983).
\bibitem{integrals} For larger chain lengths, the multi--dimensional
integrals involved become rapidly intractable.
\bibitem{privandfish} V. Privman and M. Fisher, J. Phys. A {\bf 16},
L295 (1983).
\bibitem{grassberger} P. Grassberger, Z. Phys. B {\bf 48}, 255 (1982).
\bibitem{barma} M. Barma and M. Fisher, Phys. Rev. Lett. {\bf 53}, 1935 (1984).
\end{references}
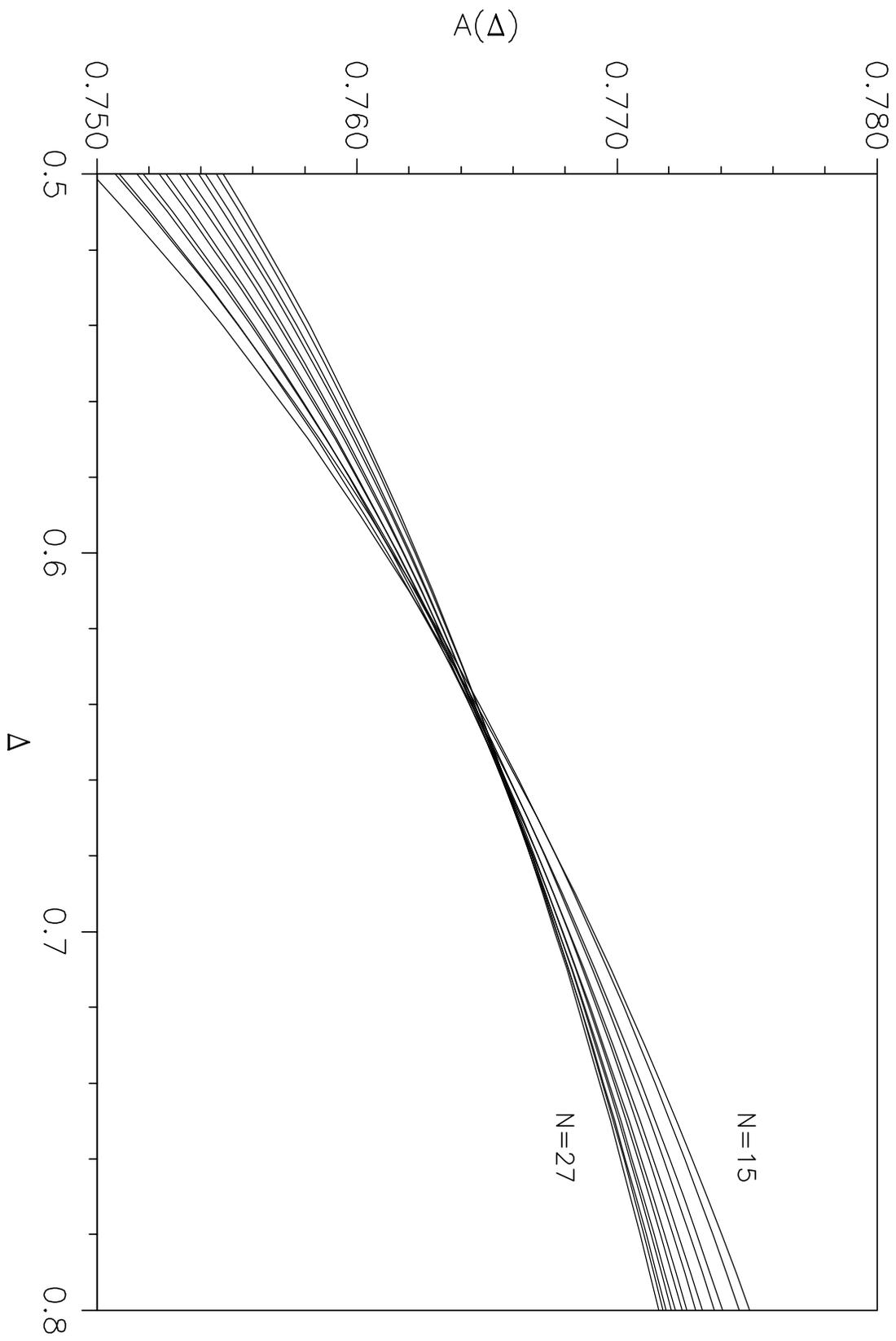
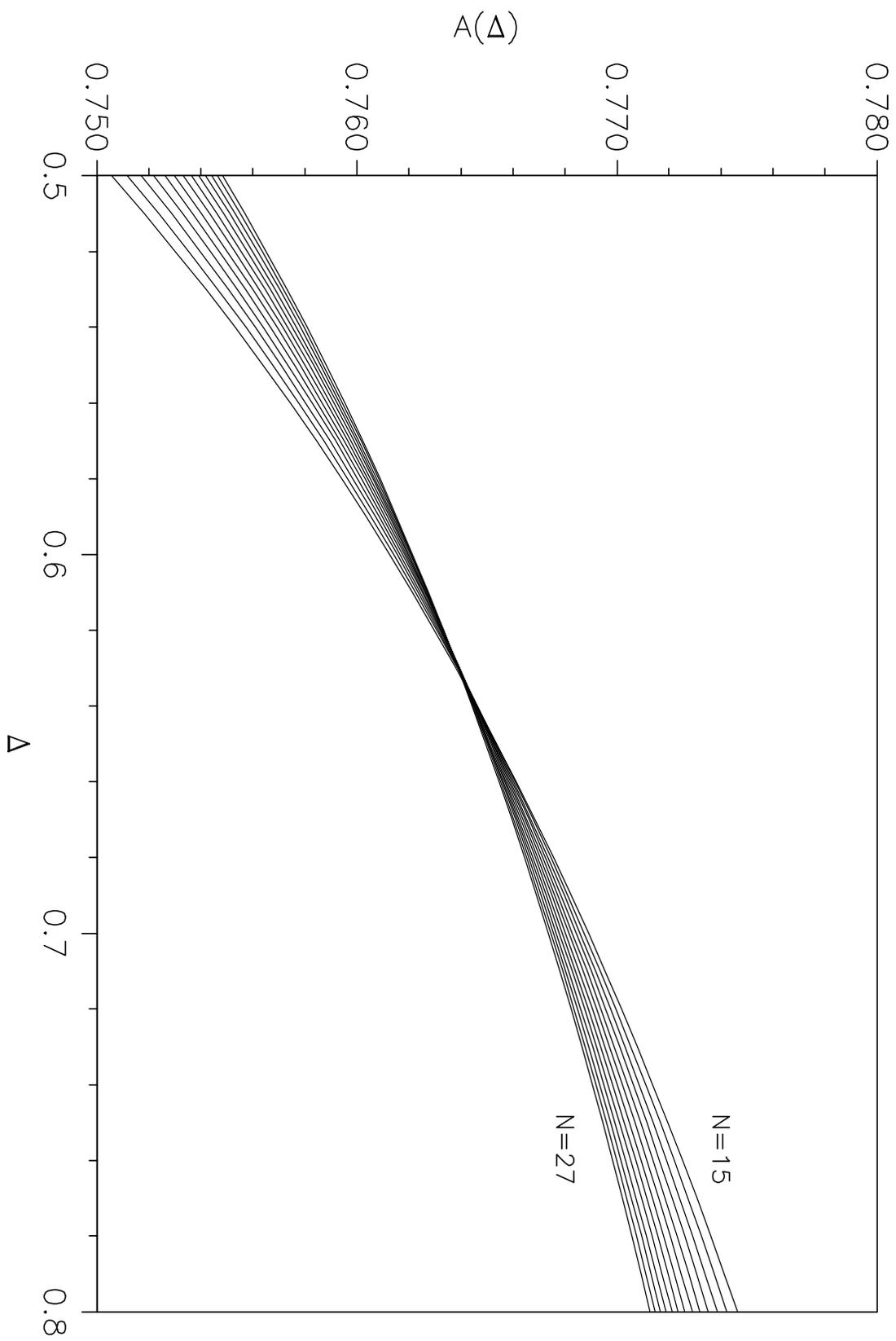
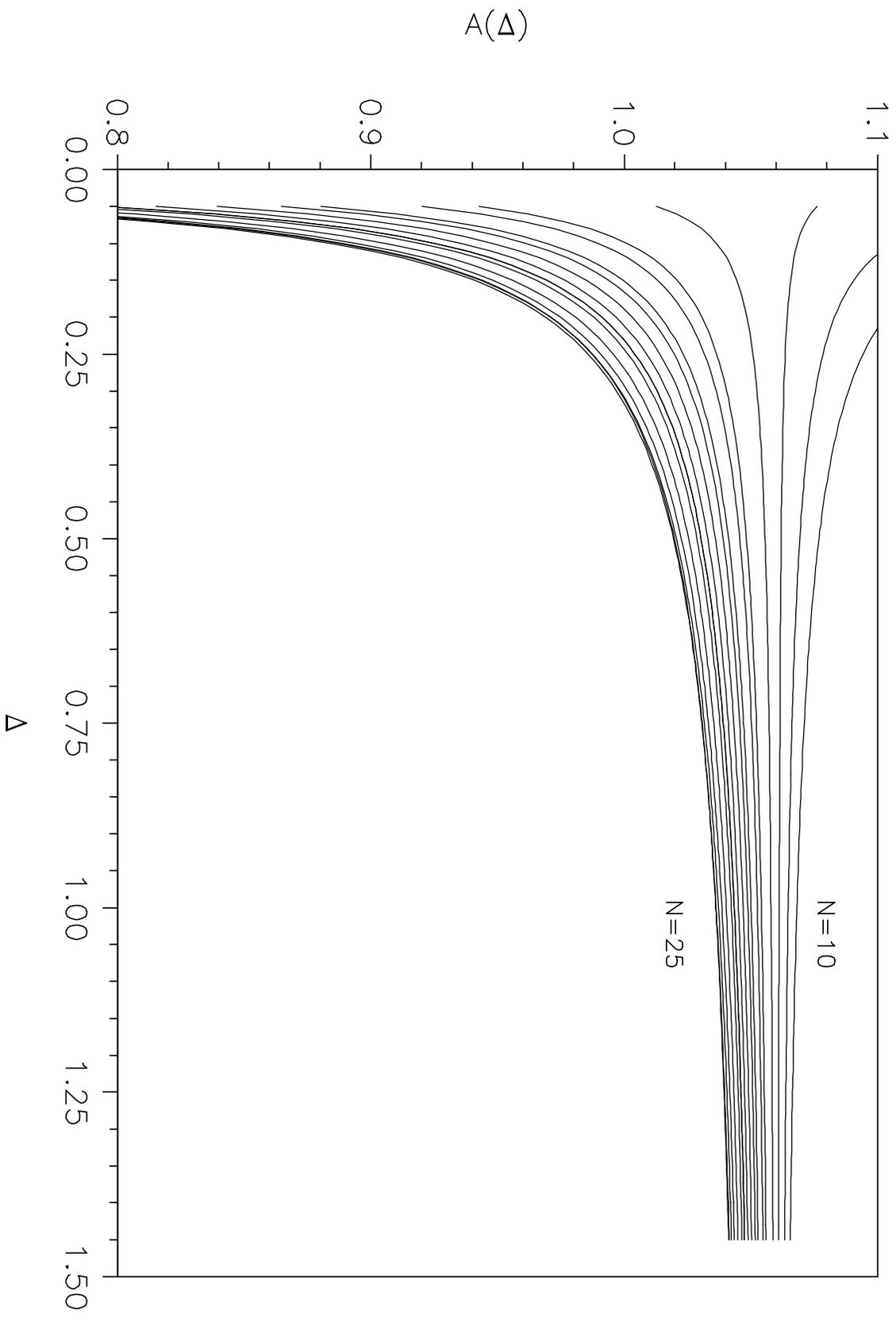
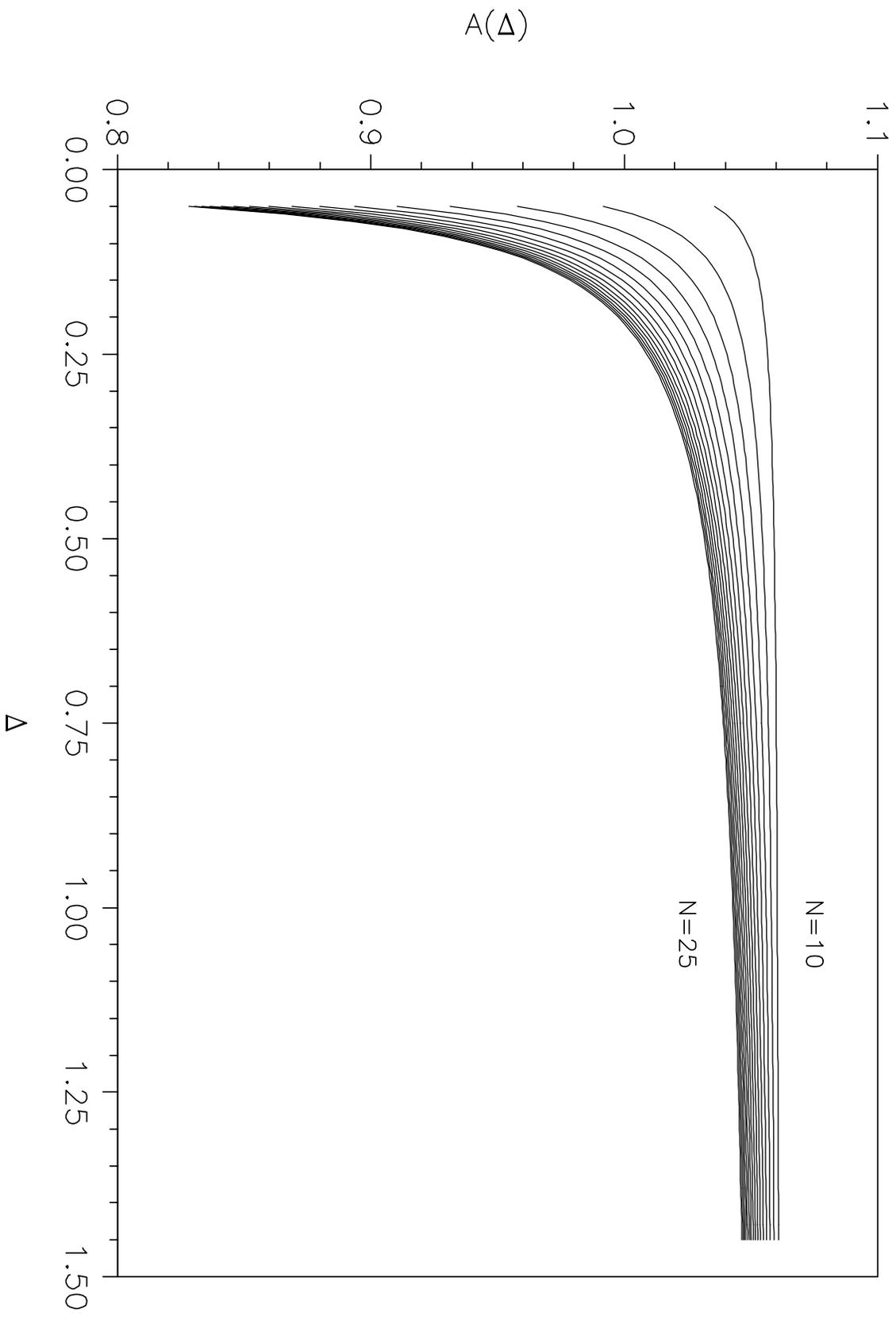
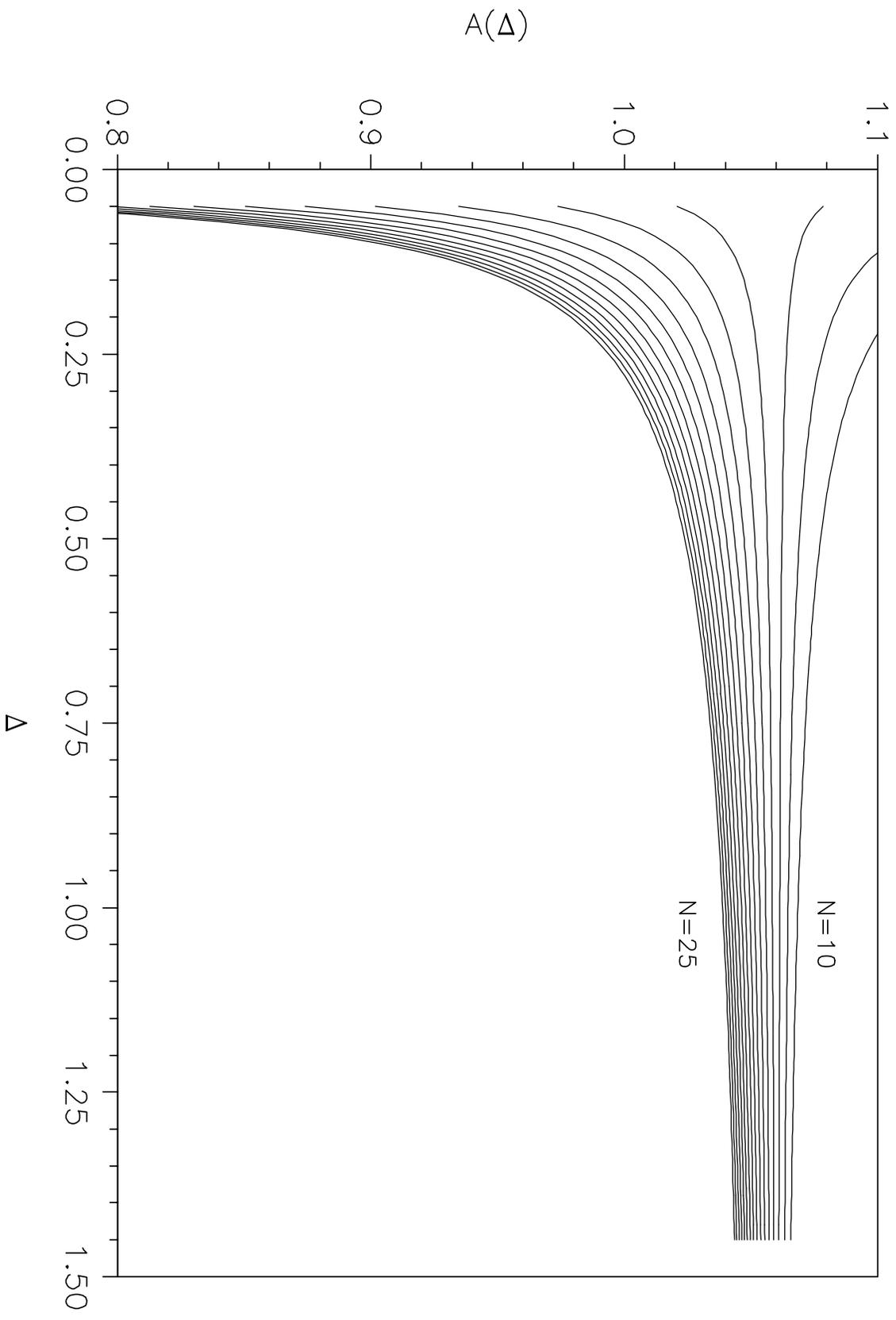
\end{document}